%% file: paper.tex
\documentclass[journal]{IEEEtran}
\usepackage[utf8]{inputenc}
\usepackage[pdftex, pdfborderstyle={/S/U/W 0}]{hyperref}
\usepackage{mwe}
\graphicspath{{./img/} }
\usepackage{tikz}
\usepackage{pgfplots}
\pgfplotsset{compat=1.8}
\usetikzlibrary{intersections,calc}

\usepackage{color,soul}

\usepackage{booktabs}
\usepackage{subcaption}
\usepackage{enumerate}

\usepackage{amsfonts}

\usepackage[colorinlistoftodos]{todonotes}

\usepackage[acronym]{glossaries}
\newacronym{phy}{PHY}{Physical}
\newacronym{mac}{MAC}{Medium Access Control}
\newacronym{ns}{NS}{Network Server}
\newacronym{gw}{GW}{Gateway}
\newacronym{ed}{ED}{End Device}
\newacronym{adr}{ADR}{Adaptive Data Rate}
\newacronym{sf}{SF}{Spreading Factor}
\newacronym{ack}{ACK}{Acknowledgment}
\newacronym{iot}{IoT}{Internet of Things}
\newacronym[plural=LPWANs,firstplural=Low Power Wide Area Networks (LPWANs)]{lpwan}{LPWAN}{Low Power Wide Area Network}
\newacronym{uplink}{UL}{uplink}
\newacronym{downlink}{DL}{downlink}
\newacronym{qos}{QoS}{Quality of Service}
\newacronym{css}{CSS}{Chirp Spread Spectrum}
\newacronym{dc}{DC}{Duty Cycle}
\newacronym{rx1}{RX1}{first receive window}
\newacronym{rx2}{RX2}{second receive window}
\newacronym{fdgw}{FDGW}{Full Duplex Gateway}
\newacronym{ttn}{TTN}{The Things Network}
\newacronym{ism}{ISM}{Industrial, Scientific, and Medical}

\author{Davide Magrin, \IEEEauthorblockN{Martina Capuzzo, and Andrea Zanella}\\
	\IEEEauthorblockA{Dept.\ of Information Engineering (DEI), University of Padova -- Via Gradenigo 6/b, 35131 Padova, Italy\\
		Email: \{magrinda, capuzzom, zanella\}@dei.unipd.it}}
\date{\today}
\title{A Thorough Study of LoRaWAN Performance \\ Under Different Parameter Settings}

\usepackage{fancyhdr}
\fancypagestyle{FirstPage}{
  \chead{This paper has been submitted to IEEE for publication. Copyright may be transferred without notice.}
}

\begin{document}

\maketitle
\thispagestyle{FirstPage} 

\begin{abstract}
  LoRaWAN is an emerging Low-Power Wide Area Network (LPWAN)
  technology, which is gaining momentum thanks to its flexibility and
  ease of deployment. Conversely to other LPWAN solutions, LoRaWAN
  indeed permits the configuration of several network parameters that
  affect different network performance indexes, such as energy
  efficiency, fairness, and capacity, in principle making it possible
  to adapt the network behavior to the specific requirements of the
  application scenario. Unfortunately, the complex and sometimes
  elusive interactions among the different network components make it
  rather difficult to predict the actual effect of a certain
  parameters setting, so that flexibility can turn into a stumbling
  block if not deeply understood. In this paper we shed light on such
  complex interactions, by observing and explaining the effect of
  different parameters settings in some illustrative scenarios. The
  simulation-based analysis reveals various trade-offs and highlights
  some inefficiencies in the design of the LoRaWAN standard.
  Furthermore, we show how significant performance gains can be
  obtained by wisely setting the system parameters, possibly in
  combination with some novel network management
  policies. 
\end{abstract}





\newacronym{phy}{PHY}{Physical}
\newacronym{gw}{GW}{Gateway}
\newacronym{ed}{ED}{End Device}
\newacronym{dr}{DR}{Data Rate}
\newacronym{adr}{ADR}{Adaptive Data Rate}
\newacronym{sf}{SF}{Spreading Factor}
\newacronym{ack}{ACK}{Acknowledgment}
\newacronym{iot}{IoT}{Internet of Things}
\newacronym{ns}{NS}{Network Server}
\newacronym[plural=LPWANs,firstplural=Low Power Wide Area Networks (LPWANs)]{lpwan}{LPWAN}{Low Power Wide Area Network}
\newacronym{ul}{UL}{uplink}
\newacronym{dl}{DL}{downlink}
\newacronym{ulpdr}{UL-PDR}{Uplink Packet Delivery Ratio}
\newacronym{qos}{QoS}{Quality of Service}
\newacronym{dc}{d.c.}{duty cycle}
\newacronym{rx1}{RX1}{first receive window}
\newacronym{rx2}{RX2}{second receive window}
\newacronym{ism}{ISM}{Industrial Scientific Medical}
\newacronym{per}{PER}{Packet Extraction Rate}
\newacronym{mac}{MAC}{Medium  Access Control}
\newacronym{cpsr}{CPSR}{Confirmed Packet Success Rate}
\newacronym{der}{DER}{Data Extraction Rate}

	\section{Introduction}\label{sec:intro}

	In the last few years the \gls{iot} paradigm has been attracting enormous
	interest from the scientific and industrial communities, thanks to the
	remarkable potential of the vision in which virtually every object can be
	remotely accessed and controlled through a connection to the Internet. Such a
	pervasive connectivity would enable various services in a wide array of
	scenarios. For example, cities could benefit from smart lighting control, a more
	efficient waste management and continuous infrastructure monitoring
	\cite{IoTSC}. In industrial scenarios, connected sensors can help to
	continuously monitor the production process, making it possible to quickly
	detect or even predict failures, while in the agricultural sector the widespread
	collection of environmental data, such as temperature and soil moisture, can
	improve quantity and quality of the soil production, while reducing costs.
	Health monitoring, home security and home automation are yet other examples of
	possible application scenarios~\cite{bandyopadhyay2011internet}.

	In general, the communication needs of such scenarios differ significantly from
	the classic high throughput and low delay requirements that have so-far driven
	the design of traditional communication systems. For example, long communication
	range and low energy consumption are more important than high bitrates, and
	supporting sporadic transmission of short packets from a massive number of
	devices is more important than providing stable high-throughput connections to
	few users \cite{M2M16}.


	Numerous technologies are currently rushing to close the gap opened by this
	paradigm shift. One prominent solution is provided by LoRaWAN, an open standard
	promoted by the LoRa Alliance that defines \gls{mac} and network management
	protocols on top of the Long Range (LoRa) \gls{phy} layer, which is instead proprietary of
	Semtech.\footnote{\url{https://www.semtech.com/products/wireless-rf/lora-transceivers}}

	The network topology is very simple: the radio signals transmitted by
	end-devices are received by one (or multiple) gateways, which then forward the
	packets to a \gls{ns} for further processing. The \gls{ns} is indeed in charge
	of the network management, which can be performed through a set of configurable
	parameters. In standard mode, communication is initiated by the end devices, but
	after each uplink transmission the devices can receive data packets or
	acknowledgements sent by the \gls{ns}. The LoRa chipset is designed to be very
	energy efficient and it promises to enable up to 10 years of energy autonomy for
	battery-powered devices. Furthermore, the transmission range has been proven to
	reach up to 1.5 km in urban scenarios and 15 km in rural areas \cite{IoTSC}. The \gls{phy}
	layer is based on a chirp modulation and supports multiple \glspl{sf}, which
	make it possible to trade bitrate for range. In addition, signals transmitted
	with different \glspl{sf} are almost orthogonal, thus potentially enabling
	multi-packet reception at the gateway. Another benefit is that the \gls{phy}
	operates on \gls{ism} bands in the MHz range, pushing down the deployment cost of the
	technology.

	The open \gls{mac} standard stimulates the creation of publicly
  available solutions for the back-end~\cite{thethings}, and makes
  scrutiny from the scientific community possible. In addition, the
  LoRaWAN standard offers large flexibility in the network
  configuration, which is another attractive factor. Indeed, the
  \gls{ns} can choose the \gls{sf} used by the different nodes, the
  duration of the receive windows, the transmission/reception
  channels, the priority of acknowledgement and downlink data packets,
  and so on. By properly setting these parameters it is hence possible
  to support reliable/bidirectional communications and to change the
  balance between communication reliability, delay, energy-efficiency
  and system capacity. However, while the effect of certain parameters
  settings can be predicted in simple scenarios, with a relatively low
  number of nodes, the interactions among the different mechanisms of
  the system become much more complex and less intuitive in the
  large-scale scenarios promised by the IoT paradigm.

	In this work, we leverage realistic network simulations to gain insight on the
  real performance of LoRaWAN technology in such scenarios, and show
  how even small adjustments in MAC layer parameters can significantly
  affect the system performance (e.g., the packet success ratio). By
  doing so, we highlight some inherent issues raised by the duty cycle
  limitations in European \gls{ism} bands, and propose some
  improvements to mitigate the impairments that LoRaWAN may experience
  at scale. Such simple ingenuities can help increase the number of
  devices that can be served by a single gateway, postponing the
  potential collapse of the network in overcrowded scenarios and
  reducing the network management costs created by inefficient network
  layouts.

	The rest of this paper is organized as follows. Sec.~\ref{sec:technology}
	provides a background on the technology, briefly describing the LoRa \gls{phy}
	modulation and LoRaWAN standard, pointing out their most important features and
	mechanisms. Sec.~\ref{sec:related} provides an outline of the state of the art
	on LoRaWAN simulations and performance investigations, while
	Sec.~\ref{sec:analysis} gives an overview of the simulation framework used in
	this article, describing the configurable parameters that have been considered
	and the simulation scenarios. In Sec.~\ref{sec:perf} we identify bottlenecks and
	relevant system dynamics. Furthermore, we discuss possible improvements to the
	standard and show how they can significantly improve the performance. Finally,
	Sec.~\ref{sec:conclusions} contains our conclusions and some possible avenues
	for future research.

	\section{Technology Overview}\label{sec:technology}

	This section discusses the key features of the LoRa modulation and chipset and
  provides an overview of the LoRaWAN standard, highlighting the aspects that
  play a major role in determining the network performance.

	\subsection{The LoRa modulation}\label{sec:lora-modulation}

	\begin{table}
		\caption{Main transmission features for different values of the SF
      parameters. Packet transmission (TX) time is specified for a
      \gls{phy} payload of 32~bytes, explicit header mode, code rate equal to 2,
      and 125 kHz channel bandwidth.}\label{tab:sf}
		\centering
		\begin{tabular}{ccccc}
			\toprule
			\gls{sf} & \acrshort{dr} & Data rate [kbit/s] & Sensitivity [dBm] & TX time [s] \\
			\midrule
			7  & 5 & 5.470 & -130.0 & 0.740 \\
			8  & 4 & 3.125 & -132.5 & 0.136 \\
			9  & 3 & 1.760 & -135.0 & 0.247 \\
			10 & 2 & 0.980 & -137.5 & 0.493 \\
			11 & 1 & 0.440 & -140.0 & 0.888 \\
			12 & 0 & 0.250 & -142.5 & 1.777 \\
			\bottomrule
		\end{tabular}
	\end{table}

	LoRa is a modulation scheme, patented by Semtech and based on chirp
  spread spectrum. Its design allows for long communication ranges,
  reaching 15~km in line-of-sight rural areas and 1.5~km in outdoor
  urban scenarios. The sensitivity (and, thus, the coverage) can be
  improved at the price of a lower bitrate, by changing the \gls{sf}
  parameter that takes integer values from 7 to 12. Higher \gls{sf}
  values correspond to lower transmission bitrates, but require a
  lower signal received power for correct reception, which turns into
  a longer coverage range. For each value of the SF parameter,
  Tab.~\ref{tab:sf} shows the associated Data Rate (DR) index, the
  nominal data rate, the sensitivity level and the transmission (TX)
  time.

	A key feature of the LoRa modulation is that packets employing different
  \glspl{sf} are almost orthogonal: transmissions overlapping in time and
  frequency can still be correctly decoded by the receiver, provided that the
  power of the target signal is sufficiently larger than that of
  interferers~\cite{croce, magrin2017performance}.

	\subsection{The LoRaWAN standard}\label{sec:lorawan-standard}

	The LoRaWAN standard~\cite{lorawan} defines \gls{mac} and network management
  protocols for devices using the LoRa modulation. The network topology is a
  star-of-stars, formed by three kinds of devices:
	\begin{itemize}
  \item \textit{\gls{ed}}: a peripheral node, typically a sensor or actuator
    that communicates only through the LoRa \gls{phy};
  \item \textit{\acrfull{ns}}: a centralized entity that controls the network
    parameters, forwards messages to applications and sends replies to the
    \glspl{ed} through the gateway(s);
  \item \textit{\gls{gw}}: an intermediate node that relays messages between
    \glspl{ed} and \gls{ns}.
	\end{itemize}

	\glspl{ed} and \glspl{gw} communicate using the LoRa modulation, while the
  connection between \glspl{gw} and \gls{ns} is realized using legacy IP
  technologies. Typically, the \glspl{gw} are equipped with LoRa chipset that
  allow for the parallel reception of multiple signals. Commercial LoRa radio
  chipsets feature 8 parallel reception paths (or chains), each of which can
  listen to a specific frequency and demodulate overlapping signals, exploiting
  the \textit{quasi} orthogonality of the different \glspl{sf}.

	LoRaWAN devices operate in unlicensed \gls{ism} bands, in specific frequencies
  which vary based on the regional regulation: in this work we will consider the
  European 868~MHz \gls{ism} frequency band, which the standard divides in four
  channels centered at 868.1~MHz, 868.3~MHz, 868.5~MHz and 868.625~MHz. Bearing
  in mind this caveat regarding the number of channels and the different channel
  access regulations, the conclusions drawn in this work have however broader
  interest, being valid also for other regions.

	\begin{table}
		\caption{LoRaWAN default channels and \acrshort{dc} limitations in
      Europe.}\label{tab:channels}
		\centering
		\begin{tabular}{lllc}
			\toprule
			Frequency (MHz) & Direction & \acrshort{dc} & Power limit (dBm) \\
			\midrule
			868.1   & DL, UL & 1\%  & 14 \\
			868.3   & DL, UL & 1\%  & 14 \\
			868.5   & DL, UL & 1\%  & 14 \\
			869.525 & DL     & 10\% & 27 \\
			\bottomrule
		\end{tabular}
	\end{table}

	As described in Tab.~\ref{tab:channels}, according to the LoRaWAN
  specifications the first three channels can be used both for \gls{ul} and
  \gls{dl} transmissions, while the 868.625~MHz channel is reserved for \gls{dl}
  communications. Moreover, the channels are regulated by different limitations
  on transmission power and \gls{dc}. In particular, the three bidirectional
  channels belong to the same sub-band and, hence, are collectively subject to a
  common DC limitation of 1\%, so that an \gls{ul} (resp. \gls{dl}) transmission
  in any of such channels will consume the \gls{ul} (resp. \gls{dl}) \gls{dc}
  budget of all three channels. Instead, the \gls{dl}-only channel at
  869.525~MHz belongs to a different sub-band that permits a \gls{dc} of 10\%
  and a larger transmission power. Note that the \gls{dr} parameter in
  Tab.~\ref{tab:sf} is generally used to indicate a pair of \gls{sf} and
  bandwidth values. Limiting our attention to the 125~kHz wide channels, the
  \gls{dr} is in one-to-one association with the \gls{sf}, as shown in
  Tab.~\ref{tab:sf}.

	The standard also defines three classes of \glspl{ed}, namely Class-A
  (\textit{All}), Class-B (\textit{Beacon)} and Class-C (\textit{Continuous}),
  which differ in the management of the \gls{dl} transmissions: Class-A devices
  can receive \gls{dl} packets only immediately after an \gls{ul} transmission;
  Class-B can schedule so-called ping-interval during which they can receive
  \gls{dl} packets; finally, Class-C can always receive packets, unless they are
  themselves transmitting.

	%

	In this work, we will focus on Class-A devices, which are the most popular and
  challenging, considering the more constrained communication capabilities. They
  are expected to be battery-powered and can receive only during two reception
  windows (\acrshort{rx1} and \acrshort{rx2}) that are opened, respectively, 1~s
  and 2~s after the end of each uplink transmission. The radio interface is then
  switched off till the next \gls{ul} to save energy. \acrshort{rx2} is opened
  only if no \gls{dl} message is successfully received during \acrshort{rx1}.
  According to the standard setting, the \gls{ns} can reply to an \gls{ul}
  transmission by sending a \gls{dl} packet in \acrshort{rx1}, using the same channel
  and \gls{sf} of the \gls{ul} packet, or in \acrshort{rx2}, using a dedicated
  channel (at 868.625~MHz in Europe) and \gls{sf}~12 (i.e., the lowest bitrate)
  to maximize the coverage range. These default settings can be changed by the
  \gls{ns} through appropriate \gls{mac} commands.

	\gls{ul} transmissions can either be unconfirmed or confirmed. In the first
  case, a message is transmitted only once and is not expected to be
  acknowledged by the \gls{ns}, while in the latter case, messages are
  retransmitted until an \gls{ack} packet is returned by the receiver, for a
  maximum of $m$ transmission attempts, with $m \in \{1,\;\dots\;,\;8\}$. Note
  that, setting $m = 1$ coincides with the unconfirmed case (the total number of
  transmissions can not exceed $m$ in both cases), but in case of confirmed
  message, the receiver will be required to generate an \gls{ack}.\footnote{Note
    that, from the \gls{gw} perspective, \gls{ack} packets are not
    distinguishable from any other \gls{dl} packet and, hence, are subject to
    the same rules and constraints.}

	For confirmed traffic, the LoRaWAN standard also provides an \gls{adr}
  mechanism through which the \gls{ns} can control the transmission parameters
  of the \gls{ed} to optimize the performance either of the device itself or of
  the global network.

  \section{Related Work}\label{sec:related}

	In the last years, the LoRaWAN technology has been the subject of many
  studies, which analyzed its performance and features with empirical
  measurements, mathematical analysis and simulative tools.

	Some seminal papers on LoRaWAN such as~\cite{petajajarvi2015coverage,
    wixted2016evaluation} test the coverage range and packet loss ratio by means
  of empirical measurements, but without investigating the impact of the
  parameters setting on the performance. Other works, such
  as~\cite{bor2017lora}, examine the impact of the modulation parameters on the
  single communication link between an \gls{ed} and its \gls{gw}, without
  considering more complex network configurations.

	To obtain more general results,~\cite{li20172d} uses a stochastic geometry
  model to jointly analyze interference in the time and frequency domains. It is
  observed that when implementing a packet repetition strategy, i.e.,
  transmitting each message multiple times, the failure probability
	reduces, but clearly the average throughput decreases because of the
  introduced redundancy. In~\cite{ferre2017collision} the author proposes
  closed-form expressions for collision and packet loss probabilities and, under
  the assumption of perfect orthogonality between \glspl{sf}, it is shown that
  the Poisson distributed process does not accurately model packet collisions in
  LoRaWAN. Network throughput, latency and collision rate for uplink
  transmissions are analyzed in~\cite{sorensen2017analysis} that, using queueing
  theory and considering the Aloha channel access protocol and the regulatory
  constrains in the use of the different sub-bands, points out the importance of
  a clever splitting of the traffic in the available sub-bands to improve the
  network performance. In~\cite{bankov2017mathematical} the authors present a
  mathematical model of the network performance, taking into account factors
  such as the capture effect and a realistic distribution of \glspl{sf} in the
  network. However, the model does not include some important network
  parameters, preventing the study of their effect on the network perfomance. A
  step further is made in~\cite{capuzzo2018mathematical} where the authors
  develop a model that makes it possible to consider various parameters
  configurations, such as the number of \glspl{ack} sent by the \gls{gw}, the
  \gls{sf} used for the downlink transmissions, and the \gls{dc} constraints
  imposed by the regulations. In this work, however, multiple retransmissions
  have not been considered.

	The study presented in~\cite{pop2017does} features a system-level analysis of
  LoRaWAN, and gives significant insights on bottlenecks and network behavior in
  presence of downlink traffic. However, besides pointing out some flaws in the
  design of the LoRaWAN medium access scheme, this work does not propose any way
  to improve the performance of the technology. In~\cite{abeele2017scalability},
  system-level simulations are again employed to assess the performance of
  confirmed and unconfirmed messages and show the detrimental impact of
  confirmation traffic on the overall network capacity and throughput. Here, the
  only proposed solution is the use of multiple gateways, without deeply
  investigating the specificities of the LoRaWAN standard.
  In~\cite{reynders2018lorawan} a module for the ns-3 simulator is proposed and
  used for a similar scope, comparing the single- and multi-gateway scenarios
  and the use of unconfirmed and confirmed messages. In this case, the authors
  correctly implement the \gls{gw}'s multiple reception paths, but do not take
  into account their association to a specific \gls{ul} frequency, which usually
  occurs during network setup: indeed, the number of packets that can be
  received simultaneously on a given frequency can not be greater than the
  number of reception paths that are listening on that frequency. Also in this
  case, the study only focuses on the performance analysis, without proposing
  any improvement.

	The authors in ~\cite{hauser2017proposal, slabickiadaptive} target the
  original \gls{adr} algorithm proposed by~\cite{thethings}, suggesting possible
  ameliorations. Generally, the modified algorithms yield an increase of network
  scalability, fairness among nodes, packet delivery ratio and robustness to
  variable channel conditions. In~\cite{reynders2017power}, the authors compute
  the optimal \glspl{sf} distribution to minimize the collision probability and
  propose a scheme to improve the fairness for nodes far from the station by
  optimally assigning \glspl{sf} and transmit power values to the network nodes,
  in order to reduce the packet error rate.

	In \cite{kouvelas2018employing} it is shown how the use of a
  persistent-Carrier Sense Multiple Access ($p$-CSMA) \gls{mac} protocol when
  transmitting \gls{ul} messages can improve the packet reception ratio.
  However, attention must be payed to the fact that having many \glspl{ed} that
  defer their transmission because of a low value of $p$ may lead to channel
  under-utilization. In \cite{8030484}, the authors investigate, via simulation,
  the impact of DC restrictions in LPWAN scenarios, showing that rate adaptation
  capabilities are indeed pivotal to maintain reasonable level of performance
  when the coverage range and the cell load increase. However, the effect of
  other parameters setting on the network performance is not considered.

	In this study we differ from the existing literature in that we target large
  networks with bidirectional traffic, a scenario that makes it possible to
  observe some unforeseen effects rising from the interaction of multiple nodes
  served by one single \gls{gw} and \gls{ns}. Furthermore, in our analysis we
  examine one by one the role played by the configurable network parameters, as
  detailed in Sec.\ref{sec:available-parameters}, thus highlighting some
  pitfalls that can affect the network performance. We then propose possible
  counteractions that require some small changes at the \gls{mac} layer, and we
  evaluate their effectiveness in some representative scenarios. As a side
  result, we enriched the ns-3 \texttt{lorawan} module with new functionalities.


	\section{Simulation setup and
    scenarios}\label{sec:simulation-software}\label{sec:analysis}

	The analysis carried out in this work leverages the ns-3
  \texttt{lorawan} module described in~\cite{magrin2017performance,
    capuzzo2018confirmed}. The module has been improved in order to
  support various network configurations and test
  the proposed ameliorations, with the aim of gaining a deeper
  understanding of the role played by each configurable parameter and
  identifying unforeseen behaviors. The module supports both confirmed
  and unconfirmed messages, permits the configuration of multiple
  network parameters, and implements a realistic model of the \gls{gw}
  chip, accounting for the eight available parallel reception paths.
  In the following of this section, we give more details on the
  available functionalities that will be leveraged to analyze the
  network.

	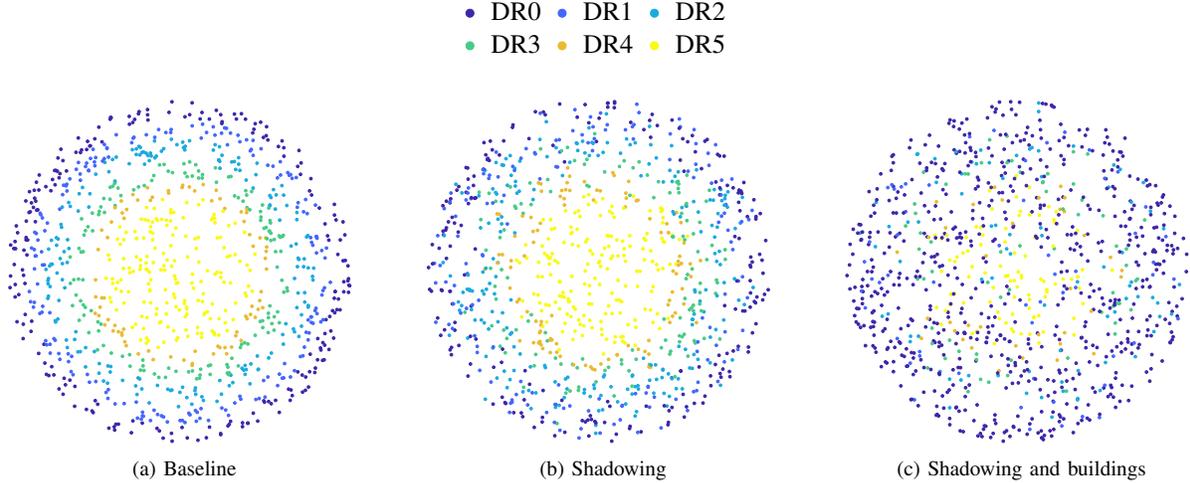
\begin{figure*}[h!]
		\centering
		\begin{subfigure}[b]{0.3\linewidth}
			\centering \input{figures/sf_plain.tex}
			\caption{Baseline}
		\end{subfigure}
		\begin{subfigure}[b]{0.3\linewidth}
			\centering \input{figures/sf_shadowing.tex}
			\caption{Shadowing}
		\end{subfigure}
		\begin{subfigure}[b]{0.3\linewidth}
			\centering \input{figures/sf_shadowing_buildings.tex}
			\caption{Shadowing and buildings}
		\end{subfigure}
		\caption{Distribution of Data Rates for different channel models.}
		\label{fig:dr_distribution}
	\end{figure*}
	\subsection{Available network settings}\label{sec:available-parameters}

	Next we give a brief introduction to the network configuration options
  that are available in the simulator and that make it possible to control the
  behavior and features of both the \gls{gw} and the \glspl{ed}.

	\begin{itemize}
  \item \textit{Gateway \gls{dc}}: in the simulator we have the opportunity of
    turning on or off the \gls{dc} restriction at the \gls{gw} to analyze its
    impact on the network performance.
  \item \textit{Transmission/reception priority}: since the \gls{gw}
    cannot receive and transmit simultaneously, this option determines
    the relative priority of transmission (TX) over reception (RX), in case of
    conflict. If priority is given to RX, then \gls{dl} packet
    transmissions will be deferred until the reception events are
    concluded (provided a suitable reception window will be open).
    Conversely, if priority is given to TX, then the reception of any
    incoming signal will be immediately interrupted to start the \gls{dl} transmission. Note that, to date, transmission prioritization
    is the only available option in commercial \glspl{gw}.
  \item \textit{Sub-band prioritization}: the LoRaWAN standard requires that
    \acrshort{rx1} is opened on the same channel where the corresponding
    \gls{ul} was received, while \acrshort{rx2} is opened on a dedicated
    \gls{dl} channel, which in Europe also features more lenient
    \gls{dc} restrictions (10\% instead of the 1\% allowed on the other channels). In the simulator, we have enabled a mode that switches this setting,
    making it possible to open \acrshort{rx1} on the dedicated \gls{dl} channel,
    and \acrshort{rx2} on the channel used for the \gls{ul} communications. The
    effect of this trick will be illustrated in Sec.~\ref{sec:ack_variations}.
  \item \textit{Acknowledgment data rate}: the LoRaWAN specifications
    recommend that \glspl{ack} transmitted on \acrshort{rx1} should
    use the same \gls{sf} for the \gls{ul} transmission, while
    transmissions on \acrshort{rx2} use the lowest available data rate
    (\gls{sf}=12). To explore other options, the simulation
    module has been modified to enable the use of higher data rates on
    both the reception windows. This setting involves a trade-off
    between robustness and efficient use of the available \gls{dc} and
    time resources. Note that such an option can actually be
    implemented in LoRaWAN through a dedicated MAC command.
  \item \textit{Number of transmission attempts}: for confirmed traffic, the
    maximum number $m$ of transmission attempts for the same message is
    configurable, and can take values in the set $\{1, 2, 4, 6, 8\}$.
  \item \textit{Full-duplex \gls{gw}}: as mentioned, currently \glspl{gw} cannot
    transmit and receive simultaneously. However, it might be interesting to
    investigate the potential performance gain obtained by implementing a
    full-duplex \gls{gw}. This functionality may be realized by co-locating two
    \glspl{gw} or combining a \gls{gw} with a simple LoRa chipset to be used for
    transmissions only, leaving the \gls{gw} free to receive messages.
    In order to test this functionality, we added a new mode to the
    \texttt{lorawan} module in the ns-3 simulator that allows for ideal full
    duplex communication.
  \item \textit{Number of reception paths}: the number $r$ of parallel
    reception paths in the \gls{gw} is a parameter that can be toggled
    in the simulator. Beside the standard value $r=8$, we also
    considered the values $r=3$ and $r=16$ to study how the parallel
    reception capabilities of the \gls{gw} can affect the overall
    system performance.

	\end{itemize}

	\subsection{Reference scenarios}\label{sec:scenario}

	We considered two main simulation scenarios. Since we are interested on the
  optimization of MAC layer parameters, we assume a single \gls{gw} serving
  multiple \glspl{ed}, which generate packets periodically, with equal period
  but random phases. Furthermore, the traffic generated by the devices can be
  either confirmed, unconfirmed, or mixed, i.e., with half of the devices
  requiring acknowledgments and the other half sending unconfirmed packets.

	In the first scenario, we assume that \glspl{ed} are randomly distributed
  within the coverage range of the \gls{gw}, and we only consider path loss.

	The second scenario consists of a more \textit{realistic urban deployment}, where
  \glspl{ed} are randomly located outside or inside buildings having different
  height and wall width, following the model described in~\cite{45.820}. Here,
  the channel propagation is affected by path loss, spatially correlated
  shadowing, and attenuation due to the presence of buildings, as described
  in~\cite{magrin2017performance}. To obtain a realistic setup, we consider the
  traffic model described in the Mobile Autonomous Reporting (MAR)
  reports~\cite{45.820}, according to which the devices send packets with
  periods that vary from 30 minutes to 24 hours. The number of devices is also
  varied to estimate the capacity (in terms of number of active devices) that
  can be supported by a \gls{gw} in a realistic scenario.

	To ease the interpretation of the results, we neglect short-term fading
  phenomena that may affect the received signal power, also considering that the
  chirp modulation is rather robust to multi-path fading.

	The effects of the channel model on the distribution of the \glspl{sf} (and,
  thus, of \glspl{dr}) can be observed in Fig.~\ref{fig:dr_distribution}, where
  dots show the position of randomly placed \glspl{ed} around the central
  \gls{gw}, while colors are used to represent the bitrate of each device, i.e.,
  its \gls{dr} value (see Tab.~\ref{tab:sf}). The bitrate is the highest
  permitted by the signal received power at the \gls{gw}, according to the
  sensitivity thresholds in Tab.~\ref{tab:sf}. Note that the rate distribution
  becomes more erratic in presence of long-term shadowing and wall attenuation
  factors that affect the propagation.

	\subsection{Performance metrics}
	\label{sec:metrics}

	A packet transmission at the \gls{phy} layer can have five possible
  outcomes:
	\begin{itemize}
  \item \textit{Success} (S): the packet is correctly received by the \gls{gw}.
  \item \textit{Lost because under sensitivity} (U): the packet arrives at the
    \gls{gw} with power lower than the sensitivity, and the \gls{gw} can not
    lock on it.
  \item \textit{Lost because of interference} (I): the packet is correctly
    locked-on by the \gls{gw}, but its reception fails because of the
    interference from overlapping packets with enough power to disrupt signals
    orthogonality.
  \item \textit{Lost because of saturated receiver} (R): the packet arrives at
    the \gls{gw} with sufficient power, but all parallel reception paths tuned
    on the packet's transmission channel are already engaged in the reception of
    other packets.
  \item \textit{Lost because of \gls{gw} transmission} (T): the packet reception
    gets disrupted by the transmission of a \gls{dl} packet (which could either
    be ongoing at the packet arrival time, or started during the packet
    reception, in case the \gls{gw} gives priority to transmission).
	\end{itemize}

	In the case of unconfirmed traffic, we label a packet as \textit{successful}
  when it is successfully received at the \gls{gw} that, in turn, forwards it to
  the \gls{ns} through a reliable connection. For confirmed traffic, we
  distinguish two cases, depending on whether the \gls{dl} packets carry
  information (e.g., the \gls{ul} packet is a query to the \gls{ns}, and the
  corresponding \gls{dl} packet is the reply), or are just an \gls{ack} used to stop
  retransmissions of the \gls{ul} packets. In the first case, the transmission
  is successful when both the \gls{ul} and the successive \gls{dl} packet are successfully
  received by the \gls{ns} and \gls{ed}, respectively, within the available
  transmission attempts. In the second case, instead, we assume that a
  transmission is successful if at least one of the generated \gls{ul} packets
  is delivered to the \gls{ns}, irrespective of whether the \gls{ack} is
  received by the device.

	%
	%
	Accordingly, we define two performance metrics:
	\begin{itemize}
  \item \gls{cpsr}: probability that both the confirmed \gls{ul}
    packet and the corresponding \gls{dl} packet are correctly
    received in one of the available transmission attempts;
  \item \gls{ulpdr}: probability that an \gls{ul} packet is correctly
    received (whether or not the \gls{ack} is requested).
	\end{itemize}


	\section{Performance evaluation}\label{sec:perf}
	In this section we first provide the baseline for our performance
  analysis considering the default settings, which reveals some issues
  with the current LoRaWAN standard. Then, we study the impact of the
  configurable parameters, and finally validate the effectiveness of
  the proposed improvements using the simulator described in
  Sec.~\ref{sec:simulation-software}.

	\subsection{Baseline performance analysis}

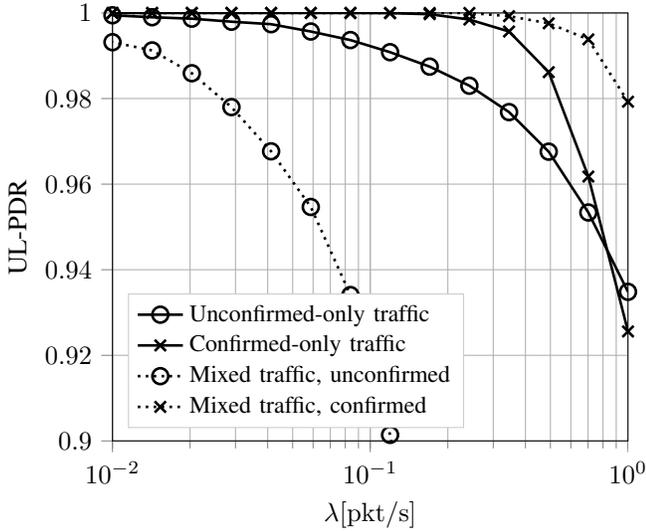
\begin{figure}
		\centering \input{figures/der.tex}
		\caption{Baseline \gls{ulpdr} performance for different kinds of traffic.}
		\label{fig:der}
	\end{figure}

	To begin with, it is interesting to compare the performance attained
  by confirmed/unconfirmed traffic in the mixed and homogeneous
  scenarios, for the same offered traffic at the application level.
  The solid lines in Fig.~\ref{fig:der} show the UL-PDR for the
  confirmed-only and unconfirmed-only cases (crossed and circle
  markers, respectively), while the dashed lines refer to the
  performance experienced by the two types of traffic sources in the
  mixed scenario. It is apparent that the mixture of confirmed and
  unconfirmed traffic sources favors the first class of sources, but
  penalizes much more severely the later, with respect to the
  corresponding homogeneous traffic cases.

	 \begin{figure}
     \centering \input{figures/phy.tex}
     \caption{PHY outcomes for traffic achieving the same \gls{ulpdr}.}
     \label{fig:phy}
   \end{figure}
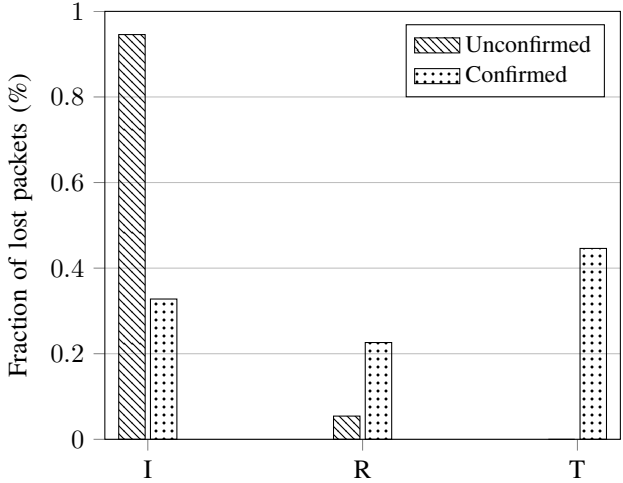

   Focusing on the homogeneous traffic scenarios, we can see that the
   use of confirmed traffic maximizes the \gls{ulpdr} index up to an
   aggregate traffic load of almost $\lambda = 0.7$~pkt/s at the
   application layer (not including retransmissions). Beyond this
   point, it is more convenient to use unconfirmed-only
   communications. The reason behind this behavior becomes apparent in
   Fig.~\ref{fig:phy}, which reports the fraction of packet losses
   caused by different events (I, R, T, see Sec.~\ref{sec:metrics})
   for the two homogeneous scenarios. The results are obtained for an
   offered traffic of $\lambda=0.7$~pkt/s, for which the \gls{ulpdr}
   is the same for both the homogeneous scenarios.
   We can observe that, with only unconfirmed traffic packet losses
   are mainly due to the interference (I) produced by multiple
   \gls{ul} transmissions. Instead, confirmed traffic (with $m = 8$),
   in addition to losses caused by interference, also suffers from
   other impairments, such as the saturation of the \gls{gw}'s
   reception paths (R), and collision with \glspl{ack} (T), which
   plays a major role among the causes of failure. Therefore,
   confirmed traffic may enhance the data collection capabilities of
   the network as long as the overall load is light, but it can yield
   significant degradation of the \gls{phy} layer performance for
   higher loads, which in turn impairs scalability.

   In the remaining of this section we will investigate the impact
   that the parameters introduced in
   Sec.~\ref{sec:available-parameters} can have on the performance
   metrics, and explore some simple precautions that can significantly
   improve both performance and fairness in LoRaWAN.

	\subsection{Gateway \gls{dc}}\label{sec:gw-dc}

	\begin{figure}[h]
		\centering
		\begin{minipage}{.48\textwidth}
			\resizebox{\textwidth}{!}{ \input{figures/cpsr.tex} }
			\caption{\gls{cpsr} of a network with only confirmed traffic sources.}
			\label{fig:cpsr}
		\end{minipage}
		\begin{minipage}{.48\textwidth}
			\vspace{2em}
		\end{minipage}
		\begin{minipage}{.48\textwidth}
			\centering \resizebox{\textwidth}{!}{ \input{figures/der_rxpriority.tex} }
			\caption{\gls{ulpdr} performance for unconfirmed, confirmed and mixed
        traffic when RX or TX is prioritized.}
			\label{fig:der_rxpriority}
		\end{minipage}
	\end{figure}

	The impact of the \gls{dc} restriction at the \gls{gw} is visible only when
  confirmed traffic is required by the \glspl{ed}. The solid line with cross
  markers in Fig.~\ref{fig:cpsr} shows the baseline \gls{cpsr} performance
  obtained in the case of only confirmed traffic. The solid line with circle
  markers, instead, gives the \gls{cpsr} that can be obtained by removing the
  \gls{dc} constraint at the \gls{gw}. Comparing the two curves it is clear
  that the \gls{dc} restriction at the \gls{gw} represents a severe bottleneck
  in terms of \gls{cpsr} since successfully received \gls{ul} packets may not be
  acknowledged by the \gls{ns} in the due time because of the \gls{dc}
  limitations of the \gls{gw}. Furthermore, the missed \glspl{ack} exacerbate
  the \gls{ul} traffic load, triggering retransmissions of otherwise
  successfully delivered \gls{ul}
  packets. 

	\subsection{Priority of transmission over
		reception}\label{sec:prior-transm-over}
	The effects of reception (RX) prioritization at the \gls{gw} have
  been investigated both in terms of \gls{cpsr} (Fig.~\ref{fig:cpsr})
  and \gls{ulpdr} (Fig.~\ref{fig:der_rxpriority}). It is worth to
  observe that RX prioritization can be implemented at the
  \gls{gw} by simply avoiding transmissions of \gls{dl} packets if at
  least one of the eight parallel receive chains is occupied.

	Fig.~\ref{fig:cpsr} shows that giving priority to RX yields a
  \gls{cpsr} loss. In fact, as $\lambda$ increases, the number of
  \gls{ul} packets that are successfully received by the \gls{gw}
  increases more rapidly than in the default case where TX is
  prioritized, and the probability that the \gls{gw} is in the
  reception state quickly approaches 1, thus preventing the \gls{gw}
  from transmitting \glspl{ack}. This, in turn, triggers packet
  re-transmissions from the devices. On the other hand, as shown in
  Fig.~\ref{fig:der_rxpriority}, in mixed traffic scenarios, the RX
  prioritization improves the performance of both confirmed and
  unconfirmed traffic sources in terms of \gls{ulpdr}. In summary,
  giving priority to RX at the \gls{gw} makes it possible to receive
  more \gls{ul} packets, but this can yield to congestion in the
  \gls{dl} channel.


	More generally, \gls{dl} packets
	could be marked by the \gls{ns} based on their importance for the \gls{ed}
  (which can either be explicitly signaled through a \gls{mac} header bit or
  inferred by the \gls{ns} based on the application that is generating the data
  flow). If \glspl{ack} are required, the \gls{dl} packet could be marked as
  prioritized over reception, and immediately sent by the \gls{gw}. If, on
  the other hand, confirmation is merely used to stop retransmissions and the
  \gls{ed} is only interested in maximizing its \gls{ulpdr}, then \glspl{ack}
  could be marked as low priority, and the \gls{gw} would send them only if
  idle.

	\subsection{ACK variations}
  \label{sec:ack_variations}

	This section analyzes the effect of two variations to the standard
  acknowledgment mechanisms, named Sub-band swapping and ACK Data Rate, that try
  to alleviate the bottleneck due to the \gls{dc} restrictions at the \gls{gw}
  and improve the system performance in terms of throughput and energy
  efficiency.

	\textit{1) Sub-band swapping:} As mentioned before, RX1 is always
  opened on the same sub-band used for the UL transmission, while RX2
  is opened on a sub-band reserved to DL transmission, whose \gls{dc}
  is $10\%$. Therefore, ACKs sent in RX1 will compete with other UL
  transmissions, generating and suffering interference, and can
  rapidly consume the $1\%$ \gls{dc} of that sub-band. We hence
  explored whether any benefit could come from swapping the sub-bands
  used for RX1 and RX2: we hence implemented a sub-band swapping
  scheme, according to which RX1 opens in the DL-reserved sub-band,
  while RX2 opens in the shared sub-band used for the UL
  transmissions.

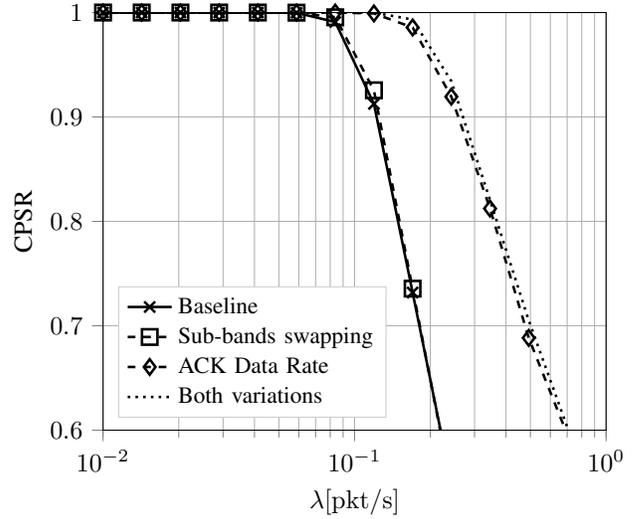
\begin{figure}[t]
		\centering \resizebox{0.48\textwidth}{!}{
      \input{figures/cpsr_improvements.tex} }
		\caption{Effect of improvements on \gls{cpsr}.}
		\label{fig:cpsr_improvements}
	\end{figure}

	\textit{2) ACK Data Rate:} By default, LoRaWAN devices use the
  highest available \gls{sf} (and thus the lowest \gls{dr}) in
  \acrshort{rx2}, in order to increase the probability that the
  downlink packet is received correctly. However, this can be
  detrimental, since longer transmission times of ACKs will rapidly
  consume the \gls{dc} budget at the GW. To study which effect is
  dominant, we implemented the ``ACK Data Rate scheme'', where all DL
  transmissions are always performed at the same DR used for the
  corresponding UL transmission. In Fig.~\ref{fig:cpsr_improvements}
  we report the \gls{cpsr} achieved by using the default setting
  (solid line with cross markers), each one of the ACK improvement
  schemes (dotted lines with square and diamond markers,
  respectively), and both the improvements together (dashed line
  without markers). We can observe that the sub-band swapping has a
  very marginal (yet positive) impact in terms of CPSR, which implies
  that the interference produced by UL transmissions on DL reception
  is not very significant. Conversely, the use of the same DR in all
  receive windows brings a significant gain in terms of CPSR over the
  baseline. We can hence conclude that the use of the lowest DR in RX2
  can severely limit the performance of the system, in particular when
  the missed reception in RX1 is not due to channel impairments, but
  rather to \gls{dc} limitations of the GW in that sub-band.

  A better strategy to provide efficient and reliable DL transmissions
  is hence to implement independent rate-adaptation strategies on all
  DL sub-bands, rather than following the very conservative policy of
  retransmitting at the basic rate to increase robustness, but at the
  cost of lower spectral efficiency.

\begin{figure}[h]
		\begin{subfigure}{0.48\textwidth}
			\centering \resizebox{\linewidth}{!}{ \input{figures/first_window.tex} }
			\caption{Average number of opened \acrshort{rx1}. \label{fig:first_window}}
		\end{subfigure}%
		\hfill%
    \vspace{2em}
		\begin{subfigure}{0.48\textwidth}
			\centering \resizebox{\linewidth}{!}{ \input{figures/second_window.tex} }
			\caption{Average number of opened \acrshort{rx2}. \label{fig:second_window}}
		\end{subfigure}%
		\hfill%
		\caption{Effects of the proposed ACK improvements on the average number of
      opened RX1 and RX2 windows.}
    \label{fig:avg_windows}
	\end{figure}
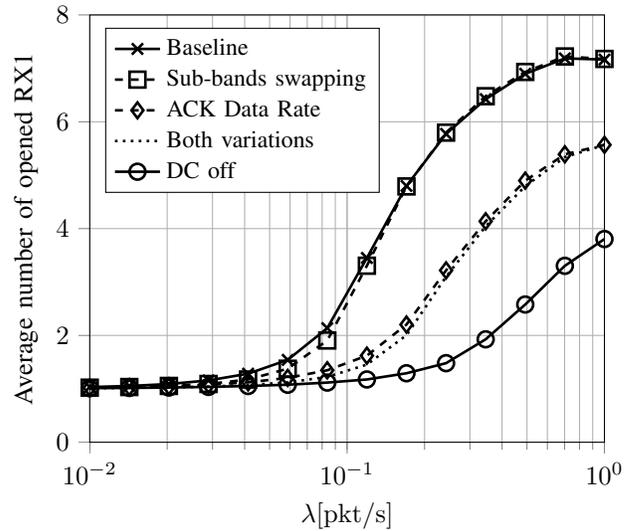
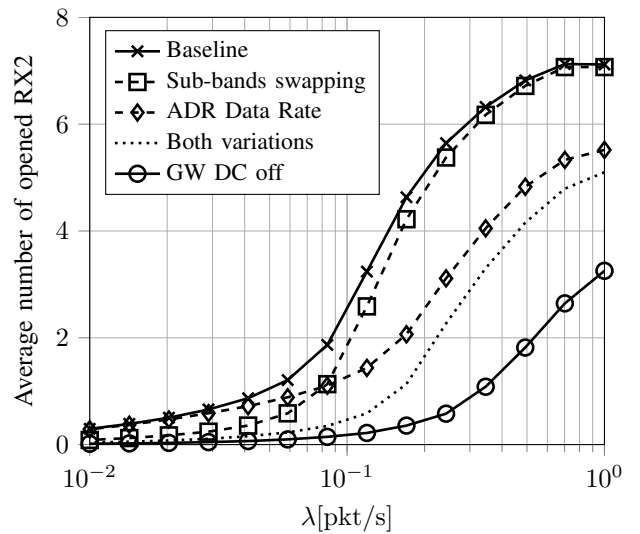

	The two ACK improvement schemes also have a positive impact on the
  energy consumption of the EDs. Indeed, the sub-band swapping
  mechanism makes it possible to return a larger number of ACKs in
  RX1, thanks to the looser \gls{dc} constraint of the DL-reserved
  sub-band, thus avoiding the need to open RX2. This effect can be
  observed in Fig.~\ref{fig:avg_windows}, which shows the average number of times RX1
  (above) and RX2 (below) are opened by the EDs, with the max number
  of retransmissions set to $m=8$. The gain, however, tends to vanish
  as the traffic increases, since both sub-bands will then be used to
  return \glspl{ack}. We can also notice that, by using the same DR in
  both receive windows we significantly reduce the average number of
  opened receive windows per transmission, also for a relatively high
  traffic. Indeed, transmitting DL packets at a higher rate
  contributes to alleviate the \gls{dc} impairments, allowing the GW
  to serve more devices. In turn, this reduces the number of
  retransmissions and, consequently, the number of \acrshort{rx1} and
  \acrshort{rx2} that need to be opened.


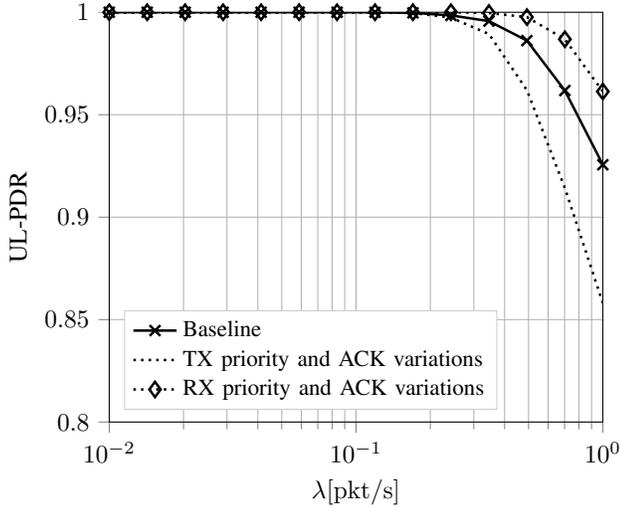
\begin{figure}[t]
		\centering \resizebox{0.48\textwidth}{!}{
      \input{figures/der_improvements.tex} }
		\caption{\gls{ulpdr} performance in case of only confirmed traffic, when ACK
      improvements and RX priority are applied.}
		\label{fig:der_improvements}
	\end{figure}

	The effect of the proposed \gls{ack} variations on the \gls{ulpdr}
  metric are depicted in Fig.~\ref{fig:der_improvements} for a network
  of only confirmed traffic sources. In this case, both sub-band
  swapping and ACK Data Rate mechanisms yield worse performance, when
  the \gls{gw} adopts the standard TX prioritization policy. This is
  easily explained if we consider the type of \gls{dl} traffic that a
  saturated network (i.e., one where the \glspl{ack} queues are always
  full) will generate when the proposed improvements are turned on and
  off: in the default case, long \gls{dl} transmissions using low data
  rates will be followed by long waiting times due to the \gls{dc}.
  During these silence periods, the \gls{gw} will be forced to listen
  to the network, resulting in an improved \gls{ulpdr} performance.
  If, on the other hand, the \gls{gw} sends short \gls{dl} packets, it
  can do so more frequently, and in turn lose more \gls{ul} packets
  because of \gls{dl} transmissions. This behavior, however, can be
  counteracted by prioritizing RX over TX: as
  Fig.~\ref{fig:der_improvements} shows, with this configuration we
  get the best of both worlds, achieving simultaneously the
  improvements on \gls{ulpdr} and the energy saving benefits obtained
  with the sub-band swapping and use of the ``ACK Data Rate'' schemes.

	One final consideration regards networks in which some devices are
  interested in the \gls{ulpdr} metric, while other need to maximize
  their \gls{cpsr}: in this case, the considerable improvement in
  \gls{cpsr} brought by the proposed acknowledgment variations would
  yield a slight loss in \gls{ulpdr}, which could be further reduced
  by implementing the dynamic transmission prioritization scheme
  proposed in Sec.~\ref{sec:prior-transm-over}.

	\subsection{Number of transmission attempts}
	Our results showed that increasing the maximum number~$m$ of transmission
  attempts improves the \gls{cpsr} by 5-10\% (though with sharply diminishing
  returns as $m$ grows larger). On the other hand, as we can see in
  Fig.~\ref{fig:der_m}, smaller values of $m$ can slightly improve the fairness
  in terms of \gls{ulpdr} in mixed traffic scenarios. In particular, at $\lambda
  = 1$ pkt/s, choosing $m = 4$ instead of $m = 8$ does not change significantly
  the \gls{ulpdr} for confirmed traffic, but yields an improvement in the
  \gls{ulpdr} of unconfirmed traffic, proving the sensitivity of the network
  performance to the setting of this parameter.

	\begin{figure}[h!]
		\centering
		\begin{minipage}{.48\textwidth}
			\resizebox{\textwidth}{!}{ \input{figures/der_m.tex} }
			\caption{\gls{ulpdr} for mixed traffic, different values of $m$.}
			\label{fig:der_m}
		\end{minipage}
		\begin{minipage}{.48\textwidth}
      \vspace{2em}
		\end{minipage}
		\begin{minipage}{.48\textwidth}
			\centering \resizebox{\textwidth}{!}{ \input{figures/der_colocation.tex} }
			\caption{Effects of Full Duplex \gls{gw} on \gls{ulpdr}.}
			\label{fig:colocation}
		\end{minipage}
	\end{figure}

	\subsection{Full-duplex gateway}
	\label{sec:colocation}

	The impact of a full-duplex \gls{gw} scheme described in
  Sec.~\ref{sec:available-parameters} is shown in
  Fig.~\ref{fig:colocation}, where \gls{ulpdr} performance is reported
  both for the standard \gls{gw} configuration and for the \gls{fdgw}.
  As expected, this solution achieves a rather marked gain in terms of
  \gls{ulpdr} performance.~\footnote{Note that, when \gls{fdgw} is
    employed, packets that are being received by the \gls{gw} are
    still lost if a transmission on that same channel is performed due
    to the strong interference.}

	\subsection{Number of available Receive Paths}

	Our simulation results (not reported here due to space constraints)
  showed that \gls{ulpdr} performance increases with the number of
  receive paths in the \gls{gw}, but with diminishing returns after 8
  reception paths, as interference still causes a relevant portion of
  locked-on packets to be lost. Having a chip with only three parallel
  receive paths, on the other hand, may enable cheaper gateways, and
  yield a slightly lower but still appreciable performance. The number
  of available receive paths, on the other hand, has no impact on the
  \gls{cpsr}, for which the bottleneck is the \gls{dl} channel due to
  the \gls{dc} constraints: by the time the additional receive paths
  can make a difference in the reception probability of \gls{ul}
  packets, \gls{dl} channels at the \gls{gw} will already be
  saturated, limiting the maximum achievable \gls{cpsr}.

	\subsection{Best configurations in a realistic scenario}

	A final simulation campaign was aimed at estimating the impact that
  the proposed variations would have on the performance of a sensor
  network deployed in a realistic urban scenario, featuring the
  channel model and \gls{sf} distribution described in
  Sec.~\ref{sec:scenario}, whose configuration is summarized in
  Tab.~\ref{tab:interarrival_times}. Note that the results have been
  plotted against the number of devices in the cell, in order to give
  an idea of the capacity gain that is achievable through a clever
  setting of the network's operational parameters.

  \begin{table}
    \caption{Interarrival times in realistic simulations.}
    \label{tab:interarrival_times}
    \centering
    \begin{tabular}{ll}
      \toprule
      Inter-arrival time & \% of devices \\
      \midrule
      1 day			& 40\%\\
      2 hours			& 40\%\\
      1 hour			& 15\%\\
      30 minutes		& 5\%\\
      \bottomrule
    \end{tabular}
  \end{table}

  \begin{table}
    \caption{Best configurations for the \gls{ulpdr} and \gls{cpsr} metrics.}
    \label{tab:best_configurations}
    \centering
    \begin{tabular}{lll}
      \toprule
      Parameter & \gls{ulpdr} & \gls{cpsr} \\
      \midrule
      \gls{ack} variations & True & True \\
      $m$ & 4 & 8 \\
      TX Prioritization & False & True \\
      \bottomrule
    \end{tabular}
  \end{table}

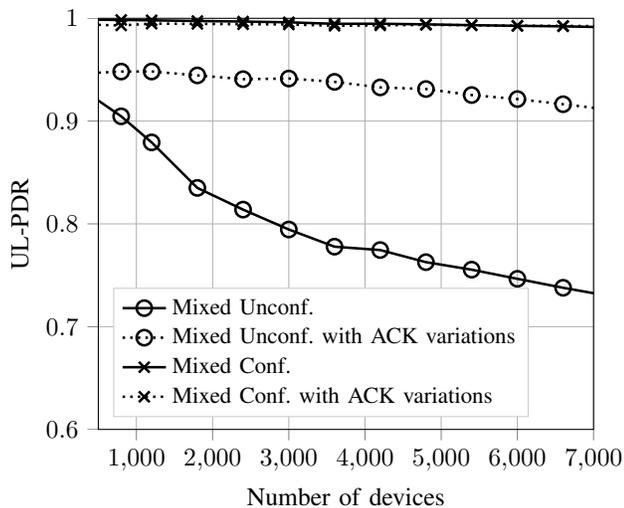
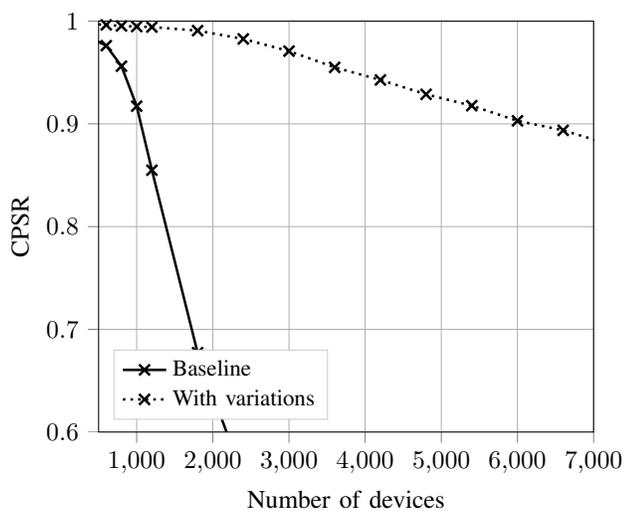
\begin{figure}
		\begin{subfigure}{0.48\textwidth}
			\resizebox{\linewidth}{!}{ \input{figures/der_realistic.tex} }
			\caption{\gls{ulpdr} performance.}
			\label{fig:realistic_der}
		\end{subfigure}%
    \vspace{2em}
		\begin{subfigure}{0.48\textwidth}
			\resizebox{\linewidth}{!}{ \input{figures/cpsr_realistic.tex} }
			\caption{\gls{cpsr} performance.}
			\label{fig:realistic_cpsr}
		\end{subfigure}%
		\hfill%
		\caption{Simulation results for a realistic scenario.}
	\end{figure}

	Fig.~\ref{fig:realistic_der} shows how the \gls{ulpdr} can improve by using
  the proposed configuration, and accommodate up to 4 times the number of
  unconfirmed devices that it would be possible to serve with standard settings.
  Similarly, Figure~\ref{fig:realistic_cpsr} shows that the number of devices
  that can be provided a \gls{cpsr} larger than 0.95 doubles when the proposed
  variations are applied.

	\subsection{Additional observations}\label{sec:other-observations}

	\subsubsection{\glspl{ed} locking on uplink packets}

	The LoRaWAN standard does not allow direct transmission between
  \glspl{ed}. Nonetheless, the simulation outcomes revealed that, when
  an \gls{ed} opens its receive window to listen for \gls{dl} packets,
  the device can actually lock onto a message sent in \gls{ul} by a
  second \gls{ed}. Experimental trials with real LoRa devices
  confirmed this incorrect behavior. This is due to the fact that the
  same preamble is used in both \gls{ul} and \gls{dl} transmissions,
  so that a receiver is not aware of the transmission source until the
  packet is completely received and inspected. At the same time, an
  \gls{ed} can also lock on a \gls{dl} message intended for another
  receiver, experiencing, thus, a waste of energy and time, as the
  packet will eventually be discarded. The problem of \glspl{ed}
  locking on \gls{ul} messages could be easily avoided by using
  different preambles for \gls{ul} and \gls{dl} transmissions: in this
  way, the receiver would completely avoid the reception of \gls{ul}
  packets and could return to sleep mode for the remaining duration of
  the \gls{ed} receive window.

	\subsubsection{Sensitivities asymmetry}

	Tab.~\ref{tab:sensitivities} shows the sensitivity requirements of
  \gls{gw} and \glspl{ed}. We can observe that the requirements for
  \glspl{ed} are more relaxed, mainly to reduce the manufacturing
  cost. However, the gap between the capabilities of the two kinds of
  device causes an asymmetric coverage range between \gls{ul} and
  \gls{dl} transmissions, and it may happen that the \gls{sf} used by
  an \gls{ed} to reach the \gls{gw} is not sufficient to correctly
  deliver the return packet to the same \gls{ed} because of its worse
  sensitivity: as an example, an \gls{ul} transmission that arrives at
  the \gls{gw} with a power of -128 dBm and SF=7 may generate a
  \gls{dl} transmission that, assuming a symmetric channel, will
  arrive with the same power at the \gls{ed}, and thus below its
  sensitivity. While such an asymmetry is not an issue when all nodes
  send unconfirmed \gls{ul} traffic, it may become a problem in case
  of confirmed traffic, since some \glspl{ed} could be prevented from
  receiving a \gls{dl} packet in the first receive window, in which
  the \gls{ns} uses the same \gls{sf} and carrier frequency of the
  \gls{ul} message, forcing the systematic opening of the second
  receive window.

	This problem can be easily mitigated at the \gls{ns} by checking
  that the reception power of the packets coming from one \gls{ed} is
  not in an interval such that the situation described above can occur
  (assuming symmetric channel). In such a case, the \gls{ns} can use
  the appropriate \gls{mac} commands to inform the device that future
  \gls{dl} transmissions in \acrshort{rx1} will use a higher \gls{sf}
  than that one used in the \gls{ul}.\footnote{The simulations in this
    article were performed by setting the \gls{ed}'s \gls{sf} based on
    the \gls{ed} sensitivities, as to avoid this asymmetry problem.}

	\begin{table}
		\caption{Sensitivity comparison}\label{tab:sensitivities}
		\centering
		\begin{tabular}{lcc}
			\toprule
			\gls{sf} & \gls{gw} Sensitivity (dBm) & \gls{ed} Sensitivity (dBm) \\
			\midrule
			7  & -130.0 & -124.0 \\
			8  & -132.5 & -127.0 \\
			9  & -135.0 & -130.0 \\
			10 & -137.5 & -133.0 \\
			11 & -140.0 & -135.0 \\
			12 & -142.5 & -137.0 \\
			\bottomrule
		\end{tabular}
	\end{table}

	\section{Conclusions}\label{sec:conclusions}

  In this article we presented a systematic analysis of the impact of
  LoRaWAN tuneable parameters on two performance indexes, namely
  UL-PDR and CPSR, which measure the effectiveness of the technology
  in collecting data from remote devices, and to support confirmed
  traffic between devices and a central controller, respectively.
  First, we observed that, with a standard settings configuration, the
  presence of confirmed traffic sources can significantly degrade the
  performance of unconfirmed traffic, due to the additional
  interference generated by \gls{dl} (\gls{ack}) transmissions.
  Considering only confirmed traffic, instead, the most critical
  factor appeared to be the \gls{dc} constraint of the \gls{gw}, which
  throttles the \gls{dl} channel that soon becomes the bottleneck of
  the system in presence of bidirectional flows. More interestingly,
  we observed that, by slightly changing the \gls{ack} procedure
  (namely, introducing the sub-band swapping and ACK Data Rate
  mechanisms) and prioritizing reception over transmission at the
  \gls{gw} (or, even better, enabling the selective prioritization of
  some \gls{dl} transmissions), it is possible to significantly
  improve the system performance in terms of packet delivery ratio,
  system capacity, energy efficiency and fairness, in particular in
  the presence of mixed traffic sources. Conversely, other system
  parameters, such as the maximum number of transmission attempts and
  the number of parallel received paths, appear to be already well
  configured and dimensioned. Overall, however, the interplay among
  the different system's tuneable knobs is often subtle and difficult
  to predict, calling for the development of efficient system
  design/configuration tools.

  \section*{Acknowledgments}

  The authors acknowledge the economical support by the POR FESR
  2014-2020 Work Program of the Veneto Region (Action 1.1.4) through
  the project No. 10066183 titled ``Sistema domotico IoT integrato ad
  elevata sicurezza informatica per smart building''.


  \bibliographystyle{IEEEtran} \bibliography{refs}

\end{document}

%% file: figures/sf_plain.tex
\definecolor{color0}{RGB}{ 61,  38,  168}
\definecolor{color1}{RGB}{ 67, 102,  253}
\definecolor{color2}{RGB}{ 27, 170,  222}
\definecolor{color3}{RGB}{ 71, 203,  134}
\definecolor{color4}{RGB}{234, 186,   48}
\definecolor{color5}{RGB}{249, 250,   20}

\begin{tikzpicture}
  \begin{axis}[
    width=7cm,
    height=7cm,
    ticks=none,
    hide axis,
    legend style={draw=none, at={(0.5, 1.1)}, anchor=north, column sep=1ex},
    legend columns=-1
    ]
    \addplot[scatter, only marks,
    point meta=explicit symbolic,
    scatter/classes={%
      0={mark=*,mark size=0.5,color0,fill=color0},
      1={mark=*,mark size=0.5,color1,fill=color1},
      2={mark=*,mark size=0.5,color2,fill=color2},
      3={mark=*,mark size=0.5,color3,fill=color3},
      4={mark=*,mark size=0.5,color4,fill=color4},
      5={mark=*,mark size=0.5,color5,fill=color5}}%
    ]%
    table [x index=0, y index=1, meta index=2]{data/sf_plain.dat};
  \end{axis}

\end{tikzpicture}

%% file: figures/sf_shadowing.tex
\definecolor{color0}{RGB}{ 61,  38,  168}
\definecolor{color1}{RGB}{ 67, 102,  253}
\definecolor{color2}{RGB}{ 27, 170,  222}
\definecolor{color3}{RGB}{ 71, 203,  134}
\definecolor{color4}{RGB}{234, 186,   48}
\definecolor{color5}{RGB}{249, 250,   20}

\begin{tikzpicture}
  \begin{axis}[
    width=7cm,
    height=7cm,
    ticks=none,
    hide axis,
    legend style={draw=none, at={(0.5, 1)}, anchor=south, column sep=1ex},
    legend image post style={scale=3},
    legend columns=3
    ]
    \addplot[scatter, only marks,
    point meta=explicit symbolic,
    scatter/classes={%
      0={mark=*,mark size=0.5,color0,fill=color0},
      1={mark=*,mark size=0.5,color1,fill=color1},
      2={mark=*,mark size=0.5,color2,fill=color2},
      3={mark=*,mark size=0.5,color3,fill=color3},
      4={mark=*,mark size=0.5,color4,fill=color4},
      5={mark=*,mark size=0.5,color5,fill=color5}}%
    ]%
    table [x index=0, y index=1, meta index=2]{data/sf_shadowing.dat};
    \legend{DR0, DR1, DR2, DR3, DR4, DR5}
  \end{axis}

\end{tikzpicture}

%% file: figures/sf_shadowing_buildings.tex
\definecolor{color0}{RGB}{ 61,  38,  168}
\definecolor{color1}{RGB}{ 67, 102,  253}
\definecolor{color2}{RGB}{ 27, 170,  222}
\definecolor{color3}{RGB}{ 71, 203,  134}
\definecolor{color4}{RGB}{234, 186,   48}
\definecolor{color5}{RGB}{249, 250,   20}

\begin{tikzpicture}
  \begin{axis}[
    width=7cm,
    height=7cm,
    ticks=none,
    hide axis,
    legend style={draw=none, at={(0.5, -0.1)}, anchor=north, column sep=1ex},
    legend columns=-1
    ]
    \addplot[scatter, only marks,
    point meta=explicit symbolic,
    scatter/classes={%
      0={mark=*,mark size=0.5,color0,fill=color0},
      1={mark=*,mark size=0.5,color1,fill=color1},
      2={mark=*,mark size=0.5,color2,fill=color2},
      3={mark=*,mark size=0.5,color3,fill=color3},
      4={mark=*,mark size=0.5,color4,fill=color4},
      5={mark=*,mark size=0.5,color5,fill=color5}}%
    ]%
    table [x index=0, y index=1, meta index=2]{data/sf_shadowing_buildings.dat};
  \end{axis}

\end{tikzpicture}

%% file: figures/der.tex
\begin{tikzpicture}

  \begin{axis}[
    legend cell align={left},
    legend entries={
      {\small{Unconfirmed-only traffic}},
      {\small{Confirmed-only traffic}},
      {\small{Mixed traffic, unconfirmed}},
      {\small{Mixed traffic, confirmed}},
    },
    legend style={at={(0.03,0.03)}, anchor=south west, draw=white!80.0!black},
    tick align=outside,
    tick pos=left,
    x tick label style = {/pgf/number format/fixed},
    x grid style={white!69.01960784313725!black},
    xlabel={$\lambda \mathrm{[pkt/s]}$},
    xmin=0.01, xmax=1,
    xmode=log,
    y grid style={white!69.01960784313725!black},
    ylabel={UL-PDR},
    ymin=0.9, ymax=1,
    grid=both,
    ]
    \addplot[mark=o, mark size=3pt, black, line width=1pt] table[x index=0,y index=1] {data/der.dat};
    \addplot[mark=x, black, mark size=3pt, line width=1pt] table[x index=0,y index=2] {data/der.dat};
    \addplot[mark=o, dotted, mark size=3pt, mark options={solid}, black, line width=1pt] table[x index=0,y index=3] {data/der.dat};
    \addplot[mark=x, dotted, black, mark size=3pt, mark options={solid}, line width=1pt] table[x index=0,y index=4] {data/der.dat};
  \end{axis}

\end{tikzpicture}

%% file: figures/phy.tex
\usetikzlibrary{patterns}
\begin{tikzpicture}

  \begin{axis}[
    legend cell align={left},
    legend entries={
      {\small{Unconfirmed}},
      {\small{Confirmed}}},
    legend pos=north east,
    tick align=outside,
    tick pos=left,
    ylabel={Fraction of lost packets (\%)},
    ymin=0, ymax=1,
    ybar,
    area legend,
    xticklabels={
      {I},
      {R},
      {T}
    },
    xtick=data,
    ymajorgrids,
    ]
    \addplot [pattern=north west lines] table[x index=0, y index=1] {data/der_normalized_phy.csv};
    \addplot [pattern=dots] table[x index=0, y index=2] {data/der_normalized_phy.csv};
  \end{axis}

\end{tikzpicture}

%% file: figures/cpsr.tex
\begin{tikzpicture}

  \begin{axis}[
    legend cell align={left},
    legend entries={
      {\small{Baseline}},
      {\small{GW DC off}},
      {\small{RX priority}},
      {\small{Double ACK}}
    },
    legend style={at={(0.03,0.03)}, anchor=south west, draw=white!80.0!black},
    tick align=outside,
    tick pos=left,
    x grid style={white!69.01960784313725!black},
    xlabel={$\lambda \mathrm{[pkt/s]}$},
    xmin=0.01, xmax=1,
    xmode=log,
    y grid style={white!69.01960784313725!black},
    ylabel={CPSR},
    ymin=0.6, ymax=1,
    grid=both,
    ]
    \addplot[mark=x, mark size=3pt, black, line width=1pt] table[x index=0,y index=2] {data/cpsr.dat};
    \addplot[mark=o, mark size=3pt, black, line width=1pt] table[x index=0,y index=1] {data/cpsr.dat};

    \addplot[dashed, mark=diamond, mark size=3pt, mark options={solid}, black, line width=1pt] table[x index=0,y index=3] {data/cpsr.dat};
  \end{axis}

\end{tikzpicture}

%% file: figures/der_rxpriority.tex
\begin{tikzpicture}

  \begin{axis}[
    legend cell align={left},
    legend entries={
      \small{Conf., TX prioritization},
      \small{Conf., RX prioritization},
      \small{Mixed Unconf., TX prioritization},
      \small{Mixed Unconf., RX prioritization},
      \small{Mixed Conf., TX prioritization},
      \small{Mixed Conf., RX prioritization},
    },
    legend style={at={(0.01,0.01)}, anchor=south west, draw=white!80.0!black},
    tick align=outside,
    tick pos=left,
    x grid style={white!69.01960784313725!black},
    xlabel={$\lambda \mathrm{[pkt/s]}$},
    xmin=0.01, xmax=1,
    xmode=log,
    y grid style={white!69.01960784313725!black},
    ylabel={UL-PDR},
    ymin=0.8, ymax=1,
    grid=both,
    ]
    \addplot[mark=diamond, black, mark size=3pt, line width=1pt] table[x index=0,y index=2] {data/der.dat};
    \addplot[mark=diamond, dashed, mark size=3pt, mark options={solid}, black, line width=1pt] table[x index=0,y index=5] {data/der.dat};

    \addplot[mark=o, dotted, mark size=3pt, mark options={solid}, black, line width=1pt] table[x index=0,y index=3] {data/der.dat};
    \addplot[mark=o, dashed, mark size=3pt, mark options={solid}, black, line width=1pt] table[x index=0,y index=6] {data/der.dat};

    \addplot[mark=x, dotted, black, mark size=3pt, mark options={solid}, line width=1pt] table[x index=0,y index=4] {data/der.dat};
    \addplot[mark=x, dashed, black, mark size=3pt, mark options={solid}, line width=1pt] table[x index=0,y index=7] {data/der.dat};
  \end{axis}

\end{tikzpicture}

%% file: figures/cpsr_improvements.tex
\begin{tikzpicture}

  \begin{axis}[
    legend cell align={left},
    legend entries={
      \small{Baseline},
      \small{Sub-bands swapping},
      \small{ACK Data Rate},
      \small{Both variations},
    },
    legend style={at={(0.03,0.03)}, anchor=south west, draw=white!80.0!black},
    tick align=outside,
    tick pos=left,
    x grid style={white!69.01960784313725!black},
    xlabel={$\lambda \mathrm{[pkt/s]}$},
    xmin=0.01, xmax=1,
    xmode=log,
    y grid style={white!69.01960784313725!black},
    ylabel={CPSR},
    ymin=0.6, ymax=1,
    grid=both,
    ]
    \addplot[mark=x, mark size=3pt, black, line width=1pt] table[x index=0,y index=2] {data/cpsr.dat};

    \addplot[dashed, mark=square, mark size=3pt, mark options={solid}, black, line width=1pt] table[x index=0,y index=8] {data/cpsr.dat};
    \addplot[dashed, mark=diamond, mark size=3pt, mark options={solid}, black, line width=1pt] table[x index=0,y index=9] {data/cpsr.dat};
    \addplot[dotted, mark size=3pt, mark options={solid}, black, line width=1pt] table[x index=0,y index=10] {data/cpsr.dat};
  \end{axis}

\end{tikzpicture}

%% file: figures/first_window.tex
\begin{tikzpicture}

  \begin{axis}[
    legend cell align={left},
    legend entries={
      \small{Baseline},
      \small{Sub-bands swapping},
      \small{ACK Data Rate},
      \small{Both variations},
      \small{DC off},
    },
    legend pos=north west,
    tick align=outside,
    tick pos=left,
    x grid style={white!69.01960784313725!black},
    xlabel={$\lambda \mathrm{[pkt/s]}$},
    xmin=0.01, xmax=1,
    xmode=log,
    y grid style={white!69.01960784313725!black},
    ylabel={Average number of opened RX1},
    ymin=0, ymax=8,
    grid=both,
    ]
    \addplot[mark=x, mark size=3pt, black, line width=1pt] table[x index=0,y index=1] {data/windows.dat};
    \addplot[dashed, mark=square, mark options={solid}, black, mark size=3pt, line width=1pt] table[x index=0,y index=2] {data/windows.dat};
    \addplot[dashed, mark=diamond, mark size=3pt, mark options={solid}, black, line width=1pt] table[x index=0,y index=3] {data/windows.dat};
    \addplot[dotted, black, mark size=3pt, mark options={solid}, line width=1pt] table[x index=0,y index=4] {data/windows.dat};
    \addplot[solid, mark=o, black, mark size=3pt, mark options={solid}, line width=1pt] table[x index=0,y index=5] {data/windows.dat};
  \end{axis}

\end{tikzpicture}

%% file: figures/second_window.tex
\begin{tikzpicture}

  \begin{axis}[
    legend cell align={left},
    legend entries={
      \small{Baseline},
      \small{Sub-bands swapping},
      \small{ADR Data Rate},
      \small{Both variations},
      \small{GW DC off},
    },
    legend pos=north west,
    tick align=outside,
    tick pos=left,
    x grid style={white!69.01960784313725!black},
    xlabel={$\lambda \mathrm{[pkt/s]}$},
    xmin=0.01, xmax=1,
    xmode=log,
    y grid style={white!69.01960784313725!black},
    ylabel={Average number of opened RX2},
    ymin=0, ymax=8,
    grid=both,
    ]
    \addplot[mark=x, mark size=3pt, black, line width=1pt] table[x index=0,y index=6] {data/windows.dat};
    \addplot[dashed, mark=square, mark options={solid}, black, mark size=3pt, line width=1pt] table[x index=0,y index=7] {data/windows.dat};
    \addplot[dashed, mark=diamond, mark size=3pt, mark options={solid}, black, line width=1pt] table[x index=0,y index=8] {data/windows.dat};
    \addplot[dotted, black, mark size=3pt, mark options={solid}, line width=1pt] table[x index=0,y index=9] {data/windows.dat};
    \addplot[solid, mark=o, black, mark size=3pt, mark options={solid}, line width=1pt] table[x index=0,y index=10] {data/windows.dat};
  \end{axis}

\end{tikzpicture}

%% file: figures/der_improvements.tex
\begin{tikzpicture}

  \begin{axis}[
    legend cell align={left},
    legend entries={
      \small{Baseline},
      \small{TX priority and ACK variations},
      \small{RX priority and ACK variations},
    },
    legend style={at={(0.03,0.03)}, anchor=south west, draw=white!80.0!black},
    tick align=outside,
    tick pos=left,
    x grid style={white!69.01960784313725!black},
    xlabel={$\lambda \mathrm{[pkt/s]}$},
    xmin=0.01, xmax=1,
    xmode=log,
    y grid style={white!69.01960784313725!black},
    ylabel={UL-PDR},
    ymin=0.8, ymax=1,
    grid=both,
    ]
    \addplot[mark=x, black, mark size=3pt, line width=1pt] table[x index=0,y index=2] {data/der.dat};
    \addplot[dotted, mark size=3pt, mark options={solid}, black, line width=1pt] table[x index=0,y index=18] {data/der.dat};
    \addplot[mark=diamond, dotted, mark size=3pt, mark options={solid}, black, line width=1pt] table[x index=0,y index=17] {data/der.dat};
  \end{axis}

\end{tikzpicture}

%% file: figures/der_m.tex
\begin{tikzpicture}

  \begin{axis}[
    legend cell align={left},
    legend entries={
      \small{Unconf., $m = 2$},
      \small{Unconf., $m = 4$},
      \small{Unconf., $m = 8$},
      \small{Conf., $m = 2$},
      \small{Conf., $m = 4$},
      \small{Conf., $m = 8$},
    },
    legend style={at={(0.01,0.01)}, anchor=south west, draw=white!80.0!black},
    tick align=outside,
    tick pos=left,
    x grid style={white!69.01960784313725!black},
    xlabel={$\lambda \mathrm{[pkt/s]}$},
    xmin=0.01, xmax=1,
    xmode=log,
    y grid style={white!69.01960784313725!black},
    ylabel={UL-PDR},
    ymin=0.6, ymax=1,
    grid=both,
    ]
    \addplot[solid, mark=o, mark size=3pt, mark options={solid}, black, line width=1pt] table[x index=0,y index=10] {data/der.dat};
    \addplot[dashed, mark=o, mark size=3pt, mark options={solid}, black, line width=1pt] table[x index=0,y index=11] {data/der.dat};
    \addplot[dotted, mark=o, mark size=3pt, mark options={solid}, black, line width=1pt] table[x index=0,y index=3] {data/der.dat};

    \addplot[solid, mark=x, black, mark size=3pt, mark options={solid}, line width=1pt] table[x index=0,y index=8] {data/der.dat};
    \addplot[dashed, mark=x, black, mark size=3pt, mark options={solid}, line width=1pt] table[x index=0,y index=9] {data/der.dat};
    \addplot[dotted, mark=x, black, mark size=3pt, mark options={solid}, line width=1pt] table[x index=0,y index=4] {data/der.dat};
  \end{axis}

\end{tikzpicture}

%% file: figures/der_colocation.tex
\begin{tikzpicture}

  \begin{axis}[
    legend cell align={left},
    legend entries={
      \small{Mixed Unconf.},
      \small{Mixed Unconf., FDGW},
      \small{Conf.},
      \small{Conf., FDGW},
      \small{Mixed Conf.},
      \small{Mixed Conf., FDGW},
    },
    legend style={at={(0.01,0.01)}, anchor=south west, draw=white!80.0!black},
    tick align=outside,
    tick pos=left,
    x grid style={white!69.01960784313725!black},
    xlabel={$\lambda \mathrm{[pkt/s]}$},
    xmin=0.01, xmax=1,
    xmode=log,
    y grid style={white!69.01960784313725!black},
    ylabel={UL-PDR},
    ymin=0.8, ymax=1,
    grid=both,
    ]
    \addplot[dotted, mark=diamond, mark size=3pt, mark options={solid}, black, line width=1pt] table[x index=0,y index=3] {data/der.dat};
    \addplot[dashed, mark=diamond, mark size=3pt, mark options={solid}, black, line width=1pt] table[x index=0,y index=12] {data/der.dat};

    \addplot[solid, mark=o, black, mark size=3pt, mark options={solid}, line width=1pt] table[x index=0,y index=2] {data/der.dat};
    \addplot[dashed, mark=o, black, mark size=3pt, mark options={solid}, line width=1pt] table[x index=0,y index=14] {data/der.dat};

    \addplot[dotted, mark=x, black, mark size=3pt, mark options={solid}, line width=1pt] table[x index=0,y index=4] {data/der.dat};
    \addplot[dashed, mark=x, black, mark size=3pt, mark options={solid}, line width=1pt] table[x index=0,y index=13] {data/der.dat};

  \end{axis}

\end{tikzpicture}

%% file: figures/der_realistic.tex
\begin{tikzpicture}

  \begin{axis}[
    legend cell align={left},
    legend entries={
      {\small{Mixed Unconf.}},
      {\small{Mixed Unconf. with ACK variations}},
      {\small{Mixed Conf.}},
      {\small{Mixed Conf. with ACK variations}},
    },
    legend style={at={(0.03,0.03)}, anchor=south west, draw=white!80.0!black},
    tick align=outside,
    tick pos=left,
    x grid style={white!69.01960784313725!black},
    xlabel={Number of devices},
    xmin=500, xmax=7000,
    y grid style={white!69.01960784313725!black},
    ylabel={UL-PDR},
    ymin=0.6, ymax=1,
    grid=both,
    ]
    \addplot[solid,mark=o, black, mark size=3pt, mark options={solid}, line width=1pt] table[x index=0,y index=3] {data/der_realistic.dat};
    \addplot[dotted,mark=o, black, mark size=3pt, mark options={solid}, line width=1pt] table[x index=0,y index=4] {data/der_realistic.dat};
    \addplot[solid,mark=x, black, mark size=3pt, mark options={solid}, line width=1pt] table[x index=0,y index=1] {data/der_realistic.dat};
    \addplot[dotted, mark=x, black, mark size=3pt, mark options={solid}, line width=1pt] table[x index=0,y index=2] {data/der_realistic.dat};
  \end{axis}

\end{tikzpicture}

%% file: figures/cpsr_realistic.tex
\begin{tikzpicture}

  \begin{axis}[
    legend cell align={left},
    legend entries={
      {\small{Baseline}},
      {\small{With variations}},
    },
    legend style={at={(0.03,0.03)}, anchor=south west, draw=white!80.0!black},
    tick align=outside,
    tick pos=left,
    x grid style={white!69.01960784313725!black},
    xlabel={Number of devices},
    y grid style={white!69.01960784313725!black},
    ylabel={CPSR},
    ymin=0.6, ymax=1,
    xmin=500, xmax=7000,
    grid=both,
    ]
    \addplot[mark=x, solid, mark size=3pt, black, line width=1pt] table[x index=0,y index=1] {data/cpsr_realistic.dat};
    \addplot[mark=x, dotted, mark size=3pt, mark options={solid}, black, line width=1pt] table[x index=0,y index=2] {data/cpsr_realistic.dat};
  \end{axis}

\end{tikzpicture}